# Dynamic Superfluid Theory of Scalar Field and Comparing Investigations with Its Corresponding Theory of Quantum Mechanics

Jia-Min Yuan[1]    Yong-Chang Huang[1, 2]

1. Institute of Theoretical Physics, Beijing University of Technology, Beijing 100124, China

2. CCAST (WorldLab.), P.O. Box 8730, 100080, Beijing, China

**ABSTRACT**

The well-known Lagrangian of current superfluid systems is not relativistic covariant, this paper gives a general relativistic covariant Lagrangian of superfluid systems, and naturally finds the non-relativistic Lagrangian and its all corresponding theories after making approximations. The equation of motion obtained from the old non-relativistic Lagrangian density is not complete, it lost some important terms. The new deduced equation can be approximated to the old equations of motion, the new momentum and energy can return to the old expressions of superfluid systems under some conditions, and the energy and momentum from the general Lagrangian density is accurate, no ignoring some terms. This paper reveals that the current classical superfluid Lagrangian density is only an approximation Lagrangian density under conditions: the scalar field is inverse proportional to its complex conjugate field and the square of time derivative of scalar field logarithm approximates to zero. This paper deduces that the divergence of velocity field is zero under classic superfluid, gives two different expressions of Lagrangian, gets the same equation, and finds its solutions. Using two different Lagrangian densities, this paper obtains different forces. Therefore, this paper discovers a general, fundamental and real physics symmetry Lagrangian deducing the real physics equation of the superfluid, and further discover the symmetry breaking processes and the special conditions from the general Lagrangian to the special Lagrangian. This paper gives both superfluid theory of scalar field and comparing investigations with its corresponding theory of quantum mechanics.

# 1. Introduction

Superfluid hydrodynamics theory has a long history. In the non-relativistic setting, Landau and Tisza put forward the ideal fluid motion equations 60 years ago [1, 2]. In the early 1980s, Israel, Khalatnikov and Lebedev extended them to relativistic superfluid motion equations [3, 4, 5]. Then Carter and Khalatnikov [6, 7] as well as Son [8] refresh it. Sonner and Withers recently did a very outstanding job [9, 10], they use the Einstein gravitational equations illustrate the Landau-Tisza theory of superfluid equations of string theory by ADS / CFT correspondence to gravity for double description. They work independently of this proves the correctness and completeness of the Landau-Tisza theory.

Landau-Tisza equation with dissipation correction is the focus of our studies today. Literature [10-16] for dissipative relativistic superfluid amendments carried out a detailed study of the textbook [17-19] in the non-relativistic limit has been amended to derive a more detailed elaboration and expansion. Although they simply lists the most general dissipative correction term, due to the presence of the stress tensor, making compliance with current and Lorentz invariance Josephson effect. Facts have proved that such a theory would bring many possibilities for non-physical. Whether or not superfluid divergence entropy flow is always positive, this is an interesting hydrodynamics were very important fact, however, this requirement gives the dissipative equations of motion correction important constraints.

As we all know, in all previous studies Lagrangian most of superfluid hydrodynamics is not utilized relativistic covariant. The study of energy-momentum tensor and the non-relativistic equations of motion are also covariant, not just regret in theory, they also do not correspond to physical perfection.

In the literature [20], in which, Son and Surowka pointed out that because of the triangle abnormal, to the normal fluid (non-superfluid) the entropy flow from its canonical form. Literature [20] made it clear that the reason for this situation is that instead of using the covariant relativistic Lagrangian is not perfect. For this reason, the starting point of this article is general covariant relativistic Lagrangian. However, we are still in this area not exceeding the requirements of symmetry assumptions.

Our results are as follows. We found that, under certain conditions, we can get the superfluid Lagrangian density from a more symmetric Lagrangian density [17]. The result provides us with an intuitive proof. Thus, while maintaining parity superfluid dissipation term structure is the most common family of 14 parameters from the literature [10] described, they are based on Clark and Putterman [18,19] of 13 parameters on initiative.

When the superfluid velocity is too large, the superfluid is no longer existing. For this reason, when taking into account the superfluid experiments, it is very often that studies of dissipative corrections are particularly interested in, which is called the normal and superfluid velocity collinear limit. In this limit parity invariant superfluid dissipation is reduced from 14 to correct parameter family of five so-called parameter family, for example, Landau and Lifshitz's classic textbook [17]. When the superfluid density is set to zero, we are in the normal phase. In this limit, two unknown parameters of superfluid theory are reduced to a constant of integration, and, with the contribution of the charging current, are proportional to the magnetic field and chiral vorticity anomalies entirely determined by the triangle [20]. Under such constraints, we can get the results of references [21-23]. In the parity preserving case superfluid, references [10-16], the authors show that it is trying to use the weight of the stream [24-29] may easily resolved by extending the superfluid dissipation correction processing to derive ultra-fluid dynamics equations. Model [30] is valid near the phase transition point closer.

In this paper, we work out a relativistic covariant Lagrangian of superfluid hydrodynamics. In other words we give the symmetric form of the equations of superfluid hydrodynamics consistent with Lorentz invariance. While we work in a relativistic covariant context throughout this paper, our final results admit a covariant limit.

Arrangements of this paper are: Sect. two is Lagrangian densities and equations of motion, Sect. three investigates conserved currents and energy-momentum tensor, Sect. four is to research on continuity equation, Sect. five studies Magnus force and the solution of the equation, Sect. six is summary and conclusion.

## 2. Lagrangian densities and equations of motion

Current Lagrangian densities (e.g., Ref.[ 31]) of the superfluid are not covariant under Lorentz transformation, thus the corresponding theories are not consistent, and which are approximate theories, thus we need the superfluid Lagrangian densities that are covariant under Lorentz transformation. In order to find the differences between covariant and non-covariant theories, we first consider the non-consistent theory as follows. Here we take the natural units throughout this paper.

The Lagrangian density of the superfluid is [31]

$$\mathcal{L} = \mathrm{i}\phi^* \partial_t \phi - \frac{1}{2}\left\{|\nabla \phi|^2 + \tau |\phi|^2 + \frac{\lambda}{2}|\phi|^4\right\}. \tag{2.1}$$

Because $\phi$, $\phi^*$ are independent，we have corresponding Euler-Lagrange Equations

$$\partial_\mu \frac{\partial \mathcal{L}}{\partial \partial_\mu \phi^*} - \frac{\partial \mathcal{L}}{\partial \phi^*} = -\mathrm{i}\partial_t \phi + \frac{1}{2}\left[-\nabla^2 \phi + \tau \phi + \lambda \phi^2 \phi^*\right] = 0, \tag{2.2}$$

$$\partial_\mu \frac{\partial \mathcal{L}}{\partial \partial_\mu \phi} - \frac{\partial \mathcal{L}}{\partial \phi} = \mathrm{i}\partial_t \phi^* + \frac{1}{2}\left[-\nabla^2 \phi^* + \tau \phi^* + \lambda |\phi|^2 \phi^*\right] = 0. \tag{2.3}$$

One gets the equation of motion

$$\mathrm{i}\dot{\phi} = \frac{1}{2}\left[-\nabla^2 + \tau + \lambda |\phi|^2\right]\phi, \tag{2.4}$$

this is the equation of motion of the superfluid.

Because the Lagrangian Eq.(2.1) deducing the interesting and real physics equation (2.4) of motion of the superfluid is non-invariant under time reverse and non-covariant under Lorentz transformation, which are absolutely unacceptable for a general, fundamental ( or original ) and real physics symmetry Lagrangian, these reveal that the Lagrangian Eq.(2.1) must be a special Lagrangian under some symmetry breakings and some special conditions from the general, fundamental and real physics symmetry Lagrangian.

Therefore, people have to find the general, fundamental and real physics symmetry Lagrangian and further discover the symmetry breaking processes and the special conditions. After having done a lot of careful investigations, we discover that the general, fundamental and real physics symmetry Lagrangian is

$$\mathcal{L} = \partial_\mu \varphi^* \partial^\mu \varphi - \alpha \varphi^* \varphi - \frac{\beta}{2} \left( \varphi^* \varphi \right)^2 , \tag{2.5}$$

which will be proved in following research.

Because $\varphi$, $\varphi^*$ are independent，so we have

$$\partial_\mu \frac{\partial \mathcal{L}}{\partial \partial_\mu \varphi^*} - \frac{\partial \mathcal{L}}{\partial \varphi^*} = \left( \partial^2 + \alpha \right) \varphi + \beta \varphi^2 \varphi^* = 0 , \tag{2.6}$$

$$\partial_\mu \frac{\partial \mathcal{L}}{\partial \partial_\mu \varphi} - \frac{\partial \mathcal{L}}{\partial \varphi} = \left( \partial^2 + \alpha \right) \varphi^* + \beta \varphi^{*2} \varphi = 0 . \tag{2.7}$$

Also we get an equation of motion

$$-\ddot{\varphi} = \left[ -\nabla^2 + \alpha + \beta |\varphi|^2 \right] \varphi . \tag{2.8}$$

In order to get the equation of motion of the superfluid from this one we do the following expansion for $\phi$ in Eq.(2.8)

$$\varphi = \phi e^{-it} , \tag{2.9}$$

$$\varphi^* = \phi^* e^{it} . \tag{2.10}$$

So we deduce

$$\dot{\varphi} = \left( \dot{\phi} - i\phi \right) e^{-it} , \tag{2.11}$$

$$\dot{\varphi}^* = \left( \dot{\phi}^* + i\phi^* \right) e^{it} . \tag{2.12}$$

From Eqs.(2.9)-(2.12), we get

$$\ddot{\varphi} = \ddot{\phi} e^{-it} - e^{-it} \left( 2i\dot{\phi} + \phi \right) , \tag{2.13}$$

$$-\nabla^2 \varphi + \alpha \varphi = \left( -\nabla^2 + \alpha \right) \phi e^{-it} , \tag{2.14}$$

$$\beta |\varphi|^2 \varphi = \beta |\phi|^2 \phi e^{-it} . \tag{2.15}$$

Put Eqs.(2.13)-(2.15) into Eq.(2.8), we get

$$\frac{\ddot{\phi}}{2} - \mathrm{i}\dot{\phi} = \frac{1}{2}\left[ \nabla^2\phi - (\alpha - 1)\phi - \beta|\phi^2|\phi \right]. \tag{2.16}$$

When taking $\alpha - 1 = \tau$, $\beta = \lambda$ and $\frac{\ddot{\phi}}{2} \approx 0$, Eq.(2.13), i.e., Eq.(2.16) becomes

$$\mathrm{i}\dot{\phi} = \frac{1}{2}\left[ -\nabla^2 + \tau + \lambda|\phi^2| \right]\phi, \tag{2.17}$$

Eq. (2.17) is just Eq.(2.4). Namely, we discover that the current classical Eq.(2.17) is only an approximation expression.

The physical meanings for $\ddot{\phi} \approx 0$ are: if $\ddot{\phi}$ $no \approx 0$, no losing generality, we have a relative force $m\ddot{\phi} = f$ $no \approx 0$, this shows that the system is under a larger force $f$, works of motion particles with the force must not be small, so energy of the system will be dissipative, i.e., this system is not stable. A system like this will not be a superfluid system. So we say that $\ddot{\phi} \approx 0$ is a condition of superfluid.

On the other hand, the physical meaning of $\ddot{\phi} \approx 0$ may be a low energy approximation for the Lorentz invariant Lagrangian density and its corresponding Euler-Lagrange equation, because the energy of the system is low, then the speed is low. If the system wants to keep small speed for long, the only probability is that $\ddot{\phi}$ keeps small for long, i.e., the acceleration can only do small fluctuation near zero to keep this small speed. So $\ddot{\phi} \approx 0$ is a condition of superfluid.

The current Schrödinger equation of the superfluid is not relativistic covariant, and will be changed under time inverse transformation, then the Lagrangian, deduced from this Schrödinger equation, is not invariant and dose not satisfy the self-consistency of physics. Because the exact Lagrangian of the real superfluid system must be Lorentz invariant, the Euler-Lagrange equation ( here i.e., the Klein-Gordon equation ) obtained from the Lorentz invariant Lagrangian is covariant [32]. There is no the first order derivative of the field about time in the Klein-Gordon equation. However there is no the second order derivative of the field about time in the current Schrödinger equation. They are not similar, there are fundamental differences between them. In fact, we have to do some approximations to deduce the usual Lagrangian of the current Schrödinger equation of the superfluid from the exact and invariant Lagrangian under general Lorentz transformation, and then obtain the usual Schrödinger equation. After these approximations about Lagrangian we find that the theory of the usual Lagrangian and Schrödinger equation of the superfluid system is just a special case of the general theory in this paper, i.e., it is an approximation theory.

The above deducing processes show that under the conditions $\alpha - 1 = \tau$, $\beta = \lambda$, $\frac{\ddot{\phi}}{2} \approx 0$, then the covariant general equation Eq.(2.8) is reduced as the non-covariant equation Eq.(2.4), i.e., under the special conditions which are achieved in this paper for the first time, the equations of motion of two Lagrangian density are the same, and our equation Eq.(2.8) is general and exact.

The current Lagrangian of superfluid systems is not relativistic covariant. So we propose a general covariant Lagrangian, then equation of motion of the Lagrangian is general and more terms than the equation of motion of the old non-covariant Lagrangian. So one can say that the equation of motion obtained from the old non-covariant Lagrangian density is not complete, however, the equation of motion obtained from the general Lagrangian density is complete. And under some special conditions the general equation of motion deduced in this paper can be approximated to the old non-covariant equations of motion.

## 3. Conserved currents and energy-momentum tensor

We have

$$\delta \mathcal{L} = \left( \frac{\partial \mathcal{L}}{\partial \varphi^a} - \partial_\mu \frac{\partial \mathcal{L}}{\partial \partial_\mu \varphi^a} \right) \delta \varphi^a + \partial_\mu \left( \frac{\partial \mathcal{L}}{\partial \partial_\mu \varphi^a} \delta \varphi^a + \mathcal{L} \Delta x^\mu \right), \qquad (3.1)$$

Then we have conservation current

$$j^\mu = \frac{\partial \mathcal{L}}{\partial \partial_\mu \varphi^a} \delta \varphi^a + \mathcal{L} \Delta x^\mu$$

$$= \partial^\mu \varphi^* \delta \varphi + \partial^\mu \varphi \delta \varphi^* + \left[ \partial_{\mu'} \varphi^* \partial^{\mu'} \varphi - \alpha \varphi^* \varphi - \frac{\beta}{2} \left( \varphi^* \varphi \right)^2 \right] \Delta x^\mu \qquad (3.2)$$

For a superfluid system, because it is a stable system, i.e., without energy dissipation, the current time component relative to the energy density must be invariant with time evolution, namely, we must have

$$\partial_0 j^0 = 0, \qquad (3.3)$$

then Eq.(3.3) can be reduced as

$$\partial_i j^i = 0. \qquad (3.4)$$

Namely, the superfluid system must be the system that the flux of the superfluid density is zero, which can be viewed as superfluid condition about a general superfluid system.

On the other hand, stable property of superfluid demands

$$\partial_0 \vec{j} = 0, \qquad (3.5)$$

then using Eqs.(3.3) and (3.5), we get

$$\partial_0 \partial_0 j^0 + \partial_i \partial_0 j^i = \partial_0 \partial_0 j^0 = 0. \qquad (3.6)$$

Evidently, Eq.(3.6) automatically includes solution (3.3). Specially, when considering $j^i = \rho v^i$, using Eq.(3.4) we deduce

$$\partial_i j^i = \partial_i (\rho v^i) = \vec{v} \cdot \nabla \rho + \rho \nabla \cdot \vec{v} = 0. \qquad (3.7)$$

When returning to classic superfluid, i.e., $\dot{x}^i \partial_i \rho(x(t)) = \frac{d\rho}{dt}$, because coordinates may be time functions for classic superfluid, Eq. (3.7) can be rewritten as

$$\frac{d\rho}{dt} + \rho \nabla \cdot \vec{v} = \frac{dj^0}{dt} + j^0 \nabla \cdot \vec{v} = 0 \qquad (3.8)$$

Using Eq.(3.3) of corresponding classic superfluid, i.e., $\frac{d\rho}{dt} = \frac{dj^0}{dt} = 0$, it follow from Eq.(3.8) that

$$\nabla \cdot \vec{v} = 0 \quad , \qquad (3.9)$$

namely, we deduce that the divergence of classic velocity field $\vec{v}$ is zero, that is, the velocity field $\vec{v}(\vec{x}(t))$ is without source field.

From Eq.(3.3) and Eq.(3.4), one gets

$$j^0 = \dot{\varphi}^* \delta\varphi + \dot{\varphi}\delta\varphi^* + \left[ \partial_\mu \varphi^* \partial^{\mu'} \varphi - \alpha \varphi^* \varphi - \frac{\beta}{2} \left( \varphi^* \varphi \right)^2 \right] \Delta x^0, \tag{3.10}$$

$$j^i = \partial^i \varphi^* \delta\varphi + \partial^i \varphi \delta\varphi^* + \left[ \partial_\mu \varphi^* \partial^{\mu'} \varphi - \alpha \varphi^* \varphi - \frac{\beta}{2} \left( \varphi^* \varphi \right)^2 \right] \Delta x^i. \tag{3.11}$$

On the other hand, because the Lagrangian density $\mathcal{L} = \mathcal{L}\left( \varphi^a, \partial_\mu \varphi^a \right)$ does not contain $x$ obviously, we can get

$$\partial_\nu T^{\mu\nu} = 0, \tag{3.12}$$

where

$$T^{\mu\nu} = \frac{\partial \mathcal{L}}{\partial \partial_\nu \varphi^a} \partial^\mu \varphi^a - g^{\mu\nu} \mathcal{L}. \tag{3.13}$$

Put Eq.(2.5) into Eq.(3.13), then we have

$$T^{\mu\nu} = \partial^\nu \varphi \partial^\mu \varphi^* + \partial^\nu \varphi^* \partial^\mu \varphi - g^{\mu\nu} \left[ \partial_\mu \varphi^* \partial^{\mu'} \varphi - \alpha \varphi^* \varphi - \frac{\beta}{2} \left( \varphi^* \varphi \right)^2 \right]. \tag{3.14}$$

We can rewrite Eq.(3.12) as

$$\partial_0 T^{\mu 0} + \nabla \cdot \vec{T}^\mu = 0, \tag{3.15}$$

where $\vec{T}^\mu = \left( T^{\mu 1}, T^{\mu 2}, T^{\mu 3} \right)$. For concrete case, taking $\mu = 0$, Eq.(3.15) can be rewritten as

$$\partial_0 E + \nabla \cdot \vec{P} = 0, \tag{3.16}$$

where $E = T^{00}, P^i = T^{0i}$, which are taken as usual in quantum field theory.

Superfluid system is a stable system, which has no energy dissipation, so that its energy density must be invariant with time evolution, namely, we must have

$$\partial_0 E = 0, \tag{3.17}$$

so Eq.(3.16) can be rewritten as

$$\nabla \cdot \vec{P} = 0, \tag{3.18}$$

where $\vec{P} = \left( T^{01}, T^{02}, T^{03} \right)$. So we can say that, the flux of momentum of the superfluid system must be zero. Namely, Eq.(3.17) and Eq.(3.18) can be viewed as superfluid conditions about a general superfluid system.

From Eq.(3.17) and Eq.(3.18) one can gets the superfluid conditions as follows

$$\partial_0 \left( \dot{\phi}\dot{\phi}^* + \left| \nabla \phi \right|^2 + \alpha \left| \phi \right|^2 + \frac{\beta}{2} \left| \phi \right|^4 \right) = 0, \tag{3.19}$$

and

$$\nabla \cdot \left( -\nabla \phi \dot{\phi}^* - \nabla \phi^* \dot{\phi} \right) = 0, \tag{3.20}$$

We get different conservation currents and the energy-momentum tensor expressions from different Lagrangians (2.1 and 2.5). A superfluid system is a stable system, and without energy dissipation, so its energy density must be invariant with time evolution. Therefore, we can deduce that, the flux of momentum of the superfluid system must be zero, which can be viewed as superfluid condition about a general superfluid system. Namely, we can view Eq.(3.3) and Eq.(3.4) or Eq.(3.17) and Eq.(3.18) as superfluid conditions about a general superfluid system. We deduce that the divergence of velocity field $\vec{v}$ is zero under classic superfluid, that is, velocity field is without source field.

## 4. Continuity equation

We now deduce continuity equation.

Using Eqs.(2.9) and (2.10), we have

$$\dot{\phi}^*\dot{\phi} = \dot{\phi}^*\dot{\phi} + 2\mathrm{i} \left( \frac{\dot{\phi}\dot{\phi}^* - \dot{\phi}^*\phi}{2} \right) + \phi\dot{\phi}^* \quad, \tag{4.1}$$

$$\nabla \phi^* \cdot \nabla \phi = \nabla \phi^* \cdot \nabla \phi \quad, \tag{4.2}$$

$$\phi^*\phi = \phi^*\phi \quad, \tag{4.3}$$

and the general Lagrangian density (2.5) can be rewritten as

$$\mathcal{L} = \dot{\phi}^*\dot{\phi} - \nabla \phi^* \cdot \nabla \phi - \alpha \phi^*\phi - \frac{\beta}{2} \left( \phi^*\phi \right)^2 \quad. \tag{4.4}$$

Put Eqs.(4.1)-(4.3) into Eq.(4.4), we have

$$\mathcal{L} = \dot{\phi}^* \dot{\phi} + 2 \left\{ i \left( \frac{\dot{\phi}\phi^* - \dot{\phi}^*\phi}{2} \right) - \frac{1}{2} \left[ |\nabla\phi|^2 + (\alpha - 1)|\phi|^2 + \frac{\beta}{2}|\phi|^4 \right] \right\} . \qquad (4.5)$$

Lagrangian Eq.(4.5) is still more general, in order to compare Eq. (4.5) with the old Lagrangian density (2.1), when taking $\dot{\phi}^*\dot{\phi} \approx 0$, $\dot{\phi}^*\phi \approx -\dot{\phi}\phi^*$, $\alpha - 1 = \tau$ and $\beta = \lambda$, then we show that Lagrangian Eq. (4.5) can be reduced as

$$\mathcal{L} = 2 \left\{ i \dot{\phi}\phi^* - \frac{1}{2} \left[ |\nabla\phi|^2 + \tau|\phi|^2 + \frac{\lambda}{2}|\phi|^4 \right] \right\}. \qquad (4.6)$$

And it follows from $\dot{\phi}^*\phi \approx -\dot{\phi}\phi^*$ that

$$\partial_t \ln\phi^* + \partial_t \ln\phi \approx 0, \quad \phi^* \approx \frac{1}{\xi(\vec{r})}\phi^{-1} \qquad , \qquad (4.7)$$

where $\xi(\vec{r})$ is an arbitrary time-independent function, i.e., we discover that the complex conjugate scalar field $\phi^*$ is inverse proportional to $\phi$, which is the new physical expression.

Using Eq.(4.7), we deduce

$$\dot{\phi}^*\dot{\phi} = -\frac{1}{\xi(\vec{r})} (\frac{\partial}{\partial t} \ln\phi)^2 \approx 0, \qquad (4.8)$$

physical meaning of Eq.(4.8) is that the square of time derivative of scalar field logarithm approximates to zero, i.e., its physical meaning is that $\ln\phi$ is such slow variable about time so that $(\frac{\partial}{\partial t} \ln\phi)^2 \approx 0$. Consequently, we discover that the current classical superfluid Lagrangian density (4.6) is only an approximation Lagrangian density under two conditions: (a) the scalar field $\phi$ is inverse proportional to its complex conjugate field $\phi^*$; (b) the square of time derivative of scalar field logarithm approximates to zero.

Eq.(4.6) is just double Lagrangian (2.1). Because any constant coefficient of Lagrangian densities doesn't affect corresponding Euler-Lagrange equations, then their Euler-Lagrange equations are the same.

When $\beta = \lambda = 2C_0^2/n_0$, $\alpha - 1 = \tau = -2C_0^2$ ( because $C_0^2$ is a positive real number in usual quantum mechanics expression of the superfluid, and because the investigations about Eq.(4.5) is general, in order to get the usual quantum mechanics expression, we must have $\alpha \leq 1$, if $\alpha > 1$, its physical meaning is that the potential configuration ($V(\phi) = (\alpha - 1)|\phi|^2 + \frac{\beta}{2}|\phi|^4$) of the system will be changed and we shall be unable to return to the usual quantum mechanics expression of the superfluid, i.e., the covariant potential is more general than the potential in the usual quantum mechanics expression ), take the Planck constant 1 and unit particle mass 1 (i.e., $M = 1$) of the superfluid (i.e., return to international unit system) and put them into Eq.(4.5), we get

$$\mathcal{L} = \dot{\phi}^* \dot{\phi} + 2 \left\{ i \left( \frac{\dot{\phi}\phi^* - \dot{\phi}^*\phi}{2} \right) - \frac{1}{2}|\nabla\phi|^2 - \frac{C_0^2}{2n_0}\left(\phi^*\phi - n_0\right)^2 + \frac{C_0^2 n_0}{2} \right\}. \qquad (4.9)$$

Euler-Lagrange equations from Eq.(4.9) is

$$i\partial_t \phi(x) = \left[ -\frac{1}{2}\nabla^2 - C_0^2 + \frac{C_0^2}{n_0}\phi^*(x)\phi(x) \right]\phi(x). \qquad (4.10)$$

For a general complex function $\phi(x)$, we can always have $\phi(x) = \rho(x)e^{i\theta'(x)}$, and we can generally define $\rho(x) = \sqrt{n(x)}$, i.e.,

$$\phi(x) = \rho(x)e^{i\theta'(x)} = \sqrt{n(x)}e^{i\theta'(x)}, \phi^*(x) = \sqrt{n(x)}e^{-i\theta'(x)}, \qquad (4.11)$$

then we have

$$\dot{\phi}(x) = \frac{1}{2\sqrt{n(x)}}\partial_t n(x)e^{i\theta'(x)} + \sqrt{n(x)}e^{i\theta'(x)} \cdot i \cdot \partial_t \theta'(x), \qquad (4.12)$$

$$\dot{\phi}^*(x) = \frac{1}{2\sqrt{n(x)}}\partial_t n(x)e^{-i\theta'(x)} - \sqrt{n(x)}e^{-i\theta'(x)} \cdot i \cdot \partial_t \theta'(x), \qquad (4.13)$$

Put Eqs.(4.11-13) into Eq.(4.9), then Eq.(4.9) can be rewritten as

$$\mathcal{L} = 2n(x)\{-\partial_t\theta'(x) - \frac{1}{2}\left[\nabla\theta'(x)\right]^2$$
$$+ e_{\nabla n}(x) - e_n(x) + \frac{\left[\partial_t\theta'(x)\right]^2}{2} + \frac{\left[\partial_t n(x)\right]^2}{8n^2(x)}\}, \qquad (4.14)$$

with

- 12 -

$$e_n(x) \equiv \frac{C_0^2}{2n_0 n(x)} \Big[ \big(n(x) - n_0\big)^2 - n_0^2 \Big] \quad , \tag{4.15}$$

Eq.(4.15) is the internal energy per particle in the fluctuating condensate, for convenience to compare with usual condensed state physics, we can define

$$\partial_t \theta'(x) \equiv \partial_t \theta(x) + \theta_t^V(x), \nabla \theta'(x) \equiv \nabla \theta(x) - \vec{\theta}^V(x) \quad , \tag{4.16}$$

$$e_{\nabla n}(x) \equiv \frac{1}{8} \frac{\big[-i\nabla n(x)\big]^2}{n^2(x)} \quad . \tag{4.17}$$

Eq. (4.17) is the gradient energy of the condensate, this energy may also be written as

$$e_{\nabla n}(x) = \frac{\big[\vec{p}^{\,os}(x)\big]^2}{2} \quad , \tag{4.18}$$

where

$$\begin{aligned}
\vec{p}^{\,os}(x) \equiv \vec{v}^{\,os}(x) &= \frac{-i}{2} \frac{\nabla n(x)}{n(x)} = \frac{1}{2} \frac{\hat{\vec{p}} n(x)}{n(x)} \\
&= \frac{1}{2} \hat{\vec{p}} \ln n(x) = \hat{\vec{p}} \ln \sqrt{n(x)} = \hat{\vec{p}} \ln \rho(x)
\end{aligned} \quad , \tag{4.19}$$

Eq. (4.19) is the quantum-mechanical momentum associated with the expansion of the condensate, namely, Eq. (4.19) can be called the osmotic momentum, and the vector $\vec{v}^{\,os}(x)$ is the associated osmotic velocity.

But in Ref.[31], the Lagrangian similar to Eq.(4.14) is

$$\mathcal{L} = n(x) \left\{ -\big[\partial_t \theta(x) + \theta_t^V(x)\big] - \frac{1}{2}\big[\nabla \theta(x) - \vec{\theta}^V(x)\big]^2 - e_{\nabla n}(x) - e_n(x) \right\}, \tag{4.20}$$

Eq. (4.20) has lost two higher terms compared with Eq.(4.15).

When the particles move in an external trap potential $V(x)$, the third and fourth terms in Eq.(4.14) can be replaced by

$$e_{tot}(x) = -e_{\nabla n}(x) + e_n(x) + V(x) \quad . \tag{4.21}$$

We may conveniently choose the axial gauge of the vortex gauge field where the time component $\theta_t^V(x)$ vanishes and only the spatial part $\vec{\theta}^V(x)$ is non zero. Then the field $\theta'(x)$ in Eq.(4.16) runs from $-\infty$ to $+\infty$ rather than $-\pi$ to $\pi$ [31].

Using Eq. (4.16), one can introduce the velocity field with vortices [31]

$$\vec{v}(x) \equiv \nabla \theta'(x) = \left[ \nabla \theta(x) - \vec{\theta}^V(x) \right], \qquad (4.22)$$

so that Eq.(4.14) can be written as

$$\mathcal{L} = -2n(x) \left[ \partial_t \theta(x) + \frac{1}{2}\vec{v}^2(x) + e_{tot}(x) - \frac{\left[ \partial_t \theta(x) \right]^2}{2} - \frac{\left[ \partial_t n(x) \right]^2}{8n^2(x)} \right], \qquad (4.23)$$

the Lagrangian density is invariant under changes of $\theta'(x)$ by an additive constant $\Lambda$. According to Noether's theorem, this implies the existence of a conservative current density. We can calculate the charge and particle current densities

$$n(x) = \frac{1}{2\left[ -1 + \partial_t \theta(x) \right]} \frac{\partial \mathcal{L}}{\partial \partial_t \theta(x)}, \qquad (4.24)$$

$$\vec{j}(x) = -\frac{\partial \mathcal{L}}{\partial \nabla \theta'(x)} = 2n(x)\vec{v}(x), \qquad (4.25)$$

where we use equation $\partial_t \theta'(x) = \partial_t \theta(x)$ in Eq.(4.24), because we have chosen the axial gauge of the vortex gauge field, i.e., the time component $\theta_t^V(x)$ is zero.

By making variation on the associated action (4.23) with respect to $\theta(x)$

$$\partial_t \left[ n(x)(1 - \partial_t \theta(x)) \right] = -\nabla \cdot \left[ n(x)\vec{v}(x) \right], \qquad (4.26)$$

Eq. (4.26) is the continuity equation of hydrodynamics.

If Lagrangian density (4.5) is replaced by (2.1), one will get the following continuity equation,

$$\partial_t n(x) = -\nabla \cdot \left[ n(x)\vec{v}(x) \right], \qquad (4.27)$$

Eq.(4.27) is missing a term compared with Eq.(4.26), because its Lagrangian density is missing two terms compared with Eq.(4.23), just like

$$\mathcal{L} = -n(x) \left[ \partial_t \theta(x) + \frac{1}{2}\vec{v}^2(x) + e_{tot}(x) \right]. \qquad (4.28)$$

The general Lagrangian density (2.5) ( this paper gives ) and the old Lagrange density (2.1) of the superfluid produce different equations, and get the different continuity equations. Compared with the new general Lagrangian density (4.14), the old one (4.20) has lost two higher terms. Our expression of osmotic momentum is exactly accord with the quantum-mechanical momentum. The general Lagrangian under the given conditions can give the complete continuity equation of fluid mechanics. This paper once again illustrates that the old Lagrangian of superfluid system and its all equations can be obtained from the new general Lagrangian under the given conditions.

## 5. Magnus force and the solution of the equation

For concrete research, we now rewrite Eq.(4.23) from the general Lagrangian as

$$\mathcal{L} = -2n(x)\left[\partial_t \theta(x) + \frac{1}{2}\vec{v}^2(x) + e_{tot}(x) - e_{ex}(x)\right], \qquad (5.1)$$

where

$$e_{ex}(x) = \frac{\left[\partial_t \theta(x)\right]^2}{2} + \frac{\left[\partial_t n(x)\right]^2}{8n^2(x)}, \qquad (5.2)$$

then Euler-Lagrange equation of Eq.(5.1) with respect to $n(x)$ yields

$$\partial_t \theta(x) + \frac{1}{2}\vec{v}^2(x) + V(x) - h_{\nabla n}(x) + h_n(x) - h_{ex}(x) = 0, \qquad (5.3)$$

where

$$h_{ex}(x) \equiv \frac{\partial\left[n(x)e_{ex}(x)\right]}{\partial n(x)} - \partial_t \frac{\partial\left[n(x)e_{ex}(x)\right]}{\partial \partial_t n(x)}, \qquad (5.4)$$

from Eq.(5.2) we can see that there is no $\nabla n(x)$ in $e_{ex}(x)$, so the term

$\nabla \cdot \dfrac{\partial\left[n(x)e_{ex}(x)\right]}{\partial \nabla n(x)}$ is zero in Eq.(5.4).

The term $h_n(x)$ is the enthalpy per particle associated with the energy $e_n(x)$, it is defined by

$$h_n(x) \equiv \frac{\partial\left[n(x)e_n(x)\right]}{\partial n(x)} = e_n(x) + n(x)\frac{\partial e_n(x)}{\partial n(x)} = e_n(x) + \frac{p_n(x)}{n(x)}, \qquad (5.5)$$

because there is no $\partial_t n(x)$ and $\nabla n(x)$ in $e_n(x)$ of Eq.(4.15), the terms

$\partial_t \dfrac{\partial\left[n(x)e_n(x)\right]}{\partial \partial_t n(x)}$ and $\nabla \cdot \dfrac{\partial\left[n(x)e_n(x)\right]}{\partial \nabla n(x)}$ are zero in Eq.(5.5). In Eq.(5.5), $p_n(x)$

is the pressure due to the energy $e_n(x)$:

$$p_n(x) \equiv n^2(x)\frac{\partial}{\partial n}e_n(x) = \left(n\frac{\partial}{\partial n} - 1\right)\left[n(x)e_n(x)\right], \qquad (5.6)$$

for $e_n(x)$, and considering that $\delta n(x) \equiv n(x) - n_0$, we find

$$h_n(x) = \frac{C_0^2}{n_0}\delta n(x), \quad p_n(x) = \frac{C_0^2}{2n_0}n^2(x). \qquad (5.7)$$

The term $h_{\nabla n}(x)$ is the so-called quantum enthalpy.

$$h_{\nabla n}(x) \equiv \frac{\partial\left[n(x)e_{\nabla n}(x)\right]}{\partial n(x)} - \nabla \cdot \frac{\partial\left[n(x)e_{\nabla n}(x)\right]}{\partial \nabla n(x)}. \qquad (5.8)$$

from Eq.(4.17) we can see there is no $\partial_t n(x)$ in $e_{\nabla n}(x)$ of Eq.(4.17), so the term

$\partial_t \dfrac{\partial\left[n(x)e_{\nabla n}(x)\right]}{\partial \partial_t n(x)}$ is zero in Eq.(5.8). Eq.(5.8) is obtained from the energy density

$e_{\nabla n}(x)$ as a contribution from the Euler-Lagrange equation, and it can be rewritten as

$$h_{\nabla n}(x) = e_{\nabla n}(x) + \frac{p_{\nabla n}(x)}{n(x)}, \qquad (5.9)$$

where

$$\begin{aligned}
p_{\nabla n}(x) &= n^2(x)\left[\frac{\partial}{\partial n} - \nabla \cdot \frac{\partial}{\partial \nabla n} - \frac{1}{n}\nabla n \cdot \frac{\partial}{\partial \nabla n}\right]e_{\nabla n}(x) \\
&= \left\{n(x)\left[\frac{\partial}{\partial n} - \nabla \cdot \frac{\partial}{\partial \nabla n}\right] - 1\right\}\left[n(x)e_{\nabla n}(x)\right]
\end{aligned}, \qquad (5.10)$$

is the so-called quantum pressure.

Inserting Eq.(4.17) into Eq.(5.8) yields

$$h_{\nabla n}(x) = \frac{1}{8n(x)}\left\{2\nabla^2 n(x) - \frac{\left[\nabla n(x)\right]^2}{n(x)}\right\}, \qquad (5.11)$$

Inserting Eq.(4.17) and Eq.(5.11) into Eq.(5.9) we get

$$p_{\nabla n}(x) = \frac{1}{4}\nabla^2 n(x). \qquad (5.12)$$

Taking the gradient of Eq.(5.3) and time derivative of Eq. (4.22), we get a useful equation

$$\partial_t \vec{v}(x) + \partial_t \vec{\theta}^V + \frac{1}{2}\nabla \vec{v}^2(x) = -\nabla V(x) + \nabla h_{\nabla n}(x) - \nabla h_n(x) + \nabla h_{ex}(x). \quad (5.13)$$

We now use the vector identity

$$\frac{1}{2}\nabla \vec{v}^2(x) = \vec{v}(x) \times \left[\nabla \times \vec{v}(x)\right] + \left[\vec{v}(x) \cdot \nabla\right]\vec{v}(x), \qquad (5.14)$$

and rewrite Eq.(5.13) as

$$\partial_t \vec{v}(x) + \left[\vec{v}(x) \cdot \nabla\right]\vec{v}(x) = -\nabla V(x) + \vec{f}^V(x), \qquad (5.15)$$

where

$$\boldsymbol{f}^V(x) \equiv -\partial_t \vec{\theta}^V - \vec{v}(x) \times \left[\nabla \times \vec{v}(x)\right] + \nabla h_{\nabla n}(x) - \nabla h_n(x) + \nabla h_{ex}(x), \quad (5.16)$$

Eq. (5.16) is a general force due to the vortices. The classical contribution to the second term is the important Magnus force acting upon a rotating fluid:

$$\vec{f}^V_{Magnus}(x) \equiv -\vec{v}(x) \times \left[\nabla \times \vec{v}(x)\right]. \qquad (5.17)$$

The important observation is now that the force expression Eq.(5.16) is, in fact, zero in the superfluid, i.e.,

$$\vec{f}^V(x) = 0, \qquad (5.18)$$

Eq. (5.18) implies that the time dependence of the vortex field is driven by the Magnus force.

The force (5.16) is perfect, because it is deduced from a general perfect Lagrangian (2.5), e.g., the deduced Eq.(5.16) from Eq. (2.5) has more terms than those in Ref.[31] on the term $\nabla h_{ex}(x)$ with important physical meanings.

Consider Lagrangian density Eq.(5.1) and omit the trivial constant condensation energy density $-C_0^2 n_0$ and higher order time derivative terms of $e_{ex}(x)$ as well as external potential $V(x)$, then we get

$$\mathcal{L} = -2\left[n_0 + \delta n(x)\right]\left[\partial_t \theta(x) + \frac{1}{2}\vec{v}^2(x)\right] - \frac{1}{4}\frac{\left[\nabla \delta n(x)\right]^2}{n(x)} - \frac{C_0^2}{n_0}\left[\delta n(x)\right]^2. \quad (5.19)$$

For small $\delta n(x) \ll n_0$, when

$$\delta n(x) = \frac{n_0}{C_0^2}\frac{-1}{1 - \xi^2\nabla^2}\left[\partial_t \theta(x) + \frac{1}{2}\vec{v}^2(x)\right], \quad (5.20)$$

where

$$\xi \equiv \frac{1}{2}\frac{1}{C_0} = \frac{1}{2}\frac{C}{C_0}\lambda_M, \quad (5.21)$$

Eq.(5.19) takes extremal value. Or taking Euler-Lagrange equation of Lagrangian density (5.19) for variable $\delta n(x)$, then we get Eqs.(5.20) and (5.21). $\lambda_M = 1/C$ is the Compton wave length of the particles of mass $M = 1$.

Reinserting Eq.(5.20) into Eq.(5.19) leads to the alternative Lagrangian density

$$\begin{aligned}\mathcal{L} = &-2n_0\left[\partial_t \theta(x) + \frac{1}{2}\vec{v}^2(x)\right] \\ &+ \frac{n_0}{C_0^2}\left[\hbar\partial_t \theta(x) + \frac{1}{2}\vec{v}^2(x)\right]\frac{1}{1 - \xi^2\nabla^2}\left[\partial_t \theta(x) + \frac{1}{2}\vec{v}^2(x)\right]\end{aligned}, \quad (5.22)$$

where $\frac{1}{4}\frac{\left[\nabla \delta n(x)\right]^2}{n(x)} \ll 1$, $\frac{C_0^2}{n_0}\left[\delta n(x)\right]^2 \ll 1$ have been neglected, doing so as in Ref.[31], and because of $\delta n(x) \ll n_0$. Then we get

$$\mathcal{L} = n_0\left\{\frac{1}{C_0^2}\left[\partial_t \theta(x)\right]\frac{1}{1 - \xi^2\nabla^2}\left[\partial_t \theta(x)\right] - 2\partial_t \theta(x) - \left[\nabla \theta(x) - \vec{\theta}^V(x)\right]^2 + \frac{1}{C_0^2}\Re\right\}, \quad (5.23)$$

where

$$\Re = \partial_t \theta(x)\frac{1}{1 - \xi^2\nabla^2}\left[\frac{1}{2}\vec{v}^2(x)\right] + \frac{1}{2}\vec{v}^2(x)\frac{1}{1 - \xi^2\nabla^2}\left[\partial_t \theta(x) + \frac{1}{2}\vec{v}^2(x)\right]. \quad (5.24)$$

When neglecting 3 order derivative terms $\Re$ of $\theta(x)$ ( due to $\vec{v}(x) = \nabla \theta'(x)$), and under the long-wavelength limit, i.e. $\xi^2 \approx 0$, which are done in Ref.[31], then the Lagrangian density Eq.(5.23) leads to the equation of motion

$$\left(-\partial_t^2 + C_0^2 \nabla^2\right)\theta(x) = 0. \tag{5.25}$$

This is a Klein-Gordon equation for $\theta(x)$ which shows that the parameter $C_0$ is the propagating velocity of phase fluctuations, which form the second sound in the superfluid [31].

Because Lagrangian density (5.23) satisfies the Euler-Lagrange equation (5.25), there must be the corresponding $\partial_\mu j^\mu = 0$, where

$$j^\mu = 2n_0 \left\{ \frac{1}{C_0^2}\left[\partial_t \theta(x)\right] - 1 - \nabla\theta(x) \right\}\delta\theta(x) + \mathcal{L}\Delta x^\mu, \tag{5.26}$$

but in Ref.[31] the Lagrangian density is

$$\mathcal{L}' = \frac{n_0}{2}\left\{ \frac{1}{C_0^2}\left[\partial_t\theta(x)\right]\frac{1}{1-\xi^2\nabla^2}\left[\partial_t\theta(x)\right] - \left[\nabla\theta(x) - \vec{\theta}^V(x)\right]^2 \right\}, \tag{5.27}$$

Eq. (5.27) is just a special case of our general Lagrangian density (5.23). Therefore, the conserved current of Lagrangian density (5.27) is

$$j'^\mu = n_0 \left\{ \frac{1}{C_0^2}\left[\partial_t\theta(x)\right] - \nabla\theta(x) \right\}\delta\theta(x) + \mathcal{L}'\Delta x^\mu. \tag{5.28}$$

Thus, the special conservation current (5.28) has lost some important terms. Namely, we have got the same equation of motion from different Lagrangian densities, but their conserved currents are different. The conserved current of general Lagrangian density is complete, and the conservative current of old Lagrangian density has lost some terms, which may have important physics. And when taking some special approximations, Eq.(5.26) can get back to Eq.(5.28).

Note the remarkable fact that although the initial equation (5.3) of motion deduced under some approximations is nonrelativistic, the sound waves follow a Lorentz-invariant equation (5.25) in which the sound velocity $C_0$ is playing the role of the light velocity. If there is a potential, the velocity of second sound will no longer be a constant but depend on the position [31].

Let us solve the equation Eq.(5.25). When $\theta_n(x) = \omega_n(t)\zeta_n(\vec{x})$ we have

$$\frac{\partial_t^2 \omega_n(t)}{\omega_n(t)} = \frac{C_0^2 \nabla^2 \zeta_n(\vec{x})}{\zeta_n(\vec{x})} = \pm A_n^2 \,, \qquad (5.29)$$

there are solutions

$$\theta(x) = \sum_n \omega_n(t)\zeta_n(\vec{x}) = \sum_n \left( C_{1n} e^{A_n t} + C_{2n} e^{-A_n t} \right) \left( C_{3n} e^{\frac{A_n}{C_0} \vec{k}_n \cdot \vec{x}} + C_{4n} e^{-\frac{A_n}{C_0} \vec{k}_n \cdot \vec{x}} \right), \qquad (5.30)$$

$$\theta(x) = \sum_n \omega_n(t)\zeta_n(\vec{x}) = \sum_n \left( b_{1n} e^{iA_n t} + b_{2n} e^{-iA_n t} \right) \left( b_{3n} e^{i\frac{A_n}{C_0} \vec{k} \cdot \vec{x}} + b_{4n} e^{-i\frac{A_n}{C_0} \vec{k} \cdot \vec{x}} \right), \qquad (5.31)$$

where $\vec{k}_n$, $(n \in Z)$, are unit vectors, $C_{in}$, $b_{in}$, $(i = 1,2,3,4)$ and $A_n$ are arbitrary constants decided by initial conditions, boundary conditions, continuity conditions and physical experiments, which are usually done in usual quantum theory, therefore, we don't repeat again here. Eq.(5.30) is not accord with the fact of stable superfluid, so it can be abandoned.

Anyone can solve Eq.(4.27) using the familiar method in general quantum theory. Because of length limit of the paper, we don't repeat here.

Being deduced from a general Lagrangian, the force deduced from Eq. (2.5) has more terms than the old one, these terms have important physics. We have given different expressions of Lagrangians and got the different equations, but include the old equation as our special example, then find general solutions of $\theta(x)$. We obtain different forces using different Lagrangians. We have got the same equation of motion from different Lagrangian densities when taking some approximations, i.e., the new deduced equation is general, and their conservation currents are different. The special conservation current has lost some terms, which may have important physical meanings. All in all, the old Lagrangian density is asymmetric, which makes the calculations lose a lot of things. And when calculating with the general Lagrangian these terms will not be lost.

# 6. Summary and conclusion

This paper not only gives superfluid theory of scalar field but also shows up comparing investigations with its corresponding theory of quantum mechanics.

The current Lagrangian of superfluid systems is not relativistic covariant, we give a general relativistic covariant Lagrangian and investigate relativistic covariant Lagrangian approximations under some conditions. Equation of motion of the new Lagrangian has more terms than those in the equation of motion of the old Lagrangian. These show that the equation of motion obtained from the old Lagrangian density is not complete, the old Lagrangian density lost the second-order terms automatically. The new general Lagrangian density solves the non-complete problem, so that the general equation of motion is complete. And under some conditions the general equation of motion can be approximated to the old equation of motion.

Although we get different conservation currents and different momentum and energy expressions from two different Lagrangians, under the some certain conditions the general momentum and energy can return to the expressions of current superfluid. These further instruct that the old Lagrangian of superfluid systems can be got by the new general Lagrangian. The energy and momentum got by the old Lagrangian density have ignored some terms, i.e., the old expressions are approximate. The energy and momentum from the new general Lagrangian density are accurate, no ignoring the higher order terms. The old energy-momentum is just a special case of new deduced energy-momentum.

After doing some approximations, transformations and overall substitutions, the new general Lagrangian density and the old Lagrange density of the superfluid produce similarity, and get the same continuity equation. That is, the new Lagrangian under the given conditions can also give the continuity equation of fluid mechanics. The old Lagrangian density (4.20) is not complete than general Lagrangian density (4.14) where the extra two terms should not be ignored for no reason. We discover that the superfluid Lagrangian density (4.6) is only an approximation Lagrangian density under two conditions: (a) the scalar field $\phi$ is inverse proportional to its complex conjugate field $\phi^*$; (b) the square of time derivative of scalar field logarithm approximates to zero. We find that $\ddot{\phi} \approx 0$ is a condition of keeping stable superfluid, and give physical meanings of the superfluid conditions deduced in this paper.

We have given two different expressions of Lagrangian densities and got the same equation. And then we find solutions of $\theta(x)$. Using two different Lagrangian densities, we obtained different forces in the expressions. The two forces only have a difference of a quantity $-\nabla h_{ex}(x)$. This amount is not reflected in the old calculation from the old Lagrangian density, this is not rigorous enough. Because the old Lagrangian density is asymmetric, which makes the calculations lose a lot of things. And with the new general Lagrangian density to calculate these terms will not be lost.

One can see that all Euler-Lagrange Equations, i.e., all the expressions, deduced from relativistic invariant Lagrangian density Eq.(2.5) are covariant when no losing any term, which are key results in quantum field theory [32], but when neglecting some terms, the all relative equations will not be covariant, e.g., the all usual expressions of the superfluid system in quantum mechanics can be achieved from the all relative expressions deduced from relativistic invariant Lagrangian density when losing some relative terms, i.e., the current Schrödinger equation and the corresponding non-invariant Lagrangian density are just special cases of the general theory in this paper and we have given out the physical meanings of the neglecting terms in process of getting the usual Schrödinger equation and the corresponding non-invariant Lagrangian density.

Therefore, this paper discovers the general, fundamental and real physics symmetry Lagrangian that can finally deduce the interesting and real physics equation (2.4) of motion of the superfluid, and further discover the symmetry breaking processes and the special conditions from the general Lagrangian (2.5) to the special Lagrangian (2.1).

Consequently, using strict, complete and relativistic covariant Lagrange density, via strict derivation, this paper makes amendment for current superfluid theory. That is, this paper gives complete and strict motion equation, gradient energy, osmotic momentum and Magnus force. After these calculation in this paper, one can see that current superfluid theory be not strict, which makes some approximations and lose some important terms. However, this paper shows that these terms have important physical significances, and should not be lost. So this paper gives a new general Lagrange density and strict computing methods, in this way, these terms are kept which has been lost in the old superfluid theory, thus the general superfluid theory is complete not as before, the appearing of these terms and the superfluid conditions have important reference values for the research on superfluid and high temperature superconductor. So the working in this paper not only has very important reference values for superfluid and superconductor in theory, but also has very important guiding significances to theories and experiments about superfluid and superconductor.

**ACKNOWLEDGMENTS:** The work is supported by National Natural Science Foundation of China (No. 11275017 and No. 11173028).